\documentclass[11pt,a4paper]{article}

\usepackage{epsfig}
\usepackage{amsmath}
\usepackage{amssymb}
\usepackage{bbm}
\usepackage{verbatim}
\usepackage{bm}
\usepackage{color}
\usepackage{dsfont}
\usepackage[footnotesize]{caption}
\usepackage[subrefformat=parens,labelformat=parens]{subfig}
\usepackage{subfloat}
\usepackage{cancel}
\usepackage{cite}
\numberwithin{equation}{section}

\usepackage{booktabs}

\def\p{\phi}
\def\pb{\bar{\phi}}

\def\d2{W''}

\def\fdir{figures/}
\let\oldincludegraphics\includegraphics
\renewcommand{\includegraphics}[2][]{\oldincludegraphics[#1]{\fdir #2}}

\begin{document}

\begin{flushleft}
CPHT-RR040.0714 \\
DESY 14-132\\
July 2014
\end{flushleft}

\vskip 1cm

\begin{center}
{\Large\bf Three-form multiplet and Inflation} 

\vskip 2cm

{Emilian Dudas$^{a,b}$} \\[3mm]
{\it{
$^a$ CPhT, Ecole Polytechnique, 91128 Palaiseau Cedex, France  \\
$^b$ Deutsches Elektronen-Synchrotron DESY, 22607 Hamburg, Germany}
}
\end{center}

\vskip 1cm

\begin{abstract}
\noindent 
Most successful models of inflation in supergravity have a shift symmetry for the inflaton and contain a stabilizer field coupled to the inflaton in a particular way. We argue that the natural interpretation of the stabilizer,  from the viewpoint of the shift symmetry,  is a three-form multiplet. Its coupling to the inflaton is uniquely determined by the shift symmetry and the invariance under three-form gauge transformations and has a natural string theory interpretation.

\end{abstract}

\thispagestyle{empty}

\newpage

%

\section{Introduction and Conclusions}

Inflation \cite{guth,linde,steinhardt} is an attractive scenario for explaining the initial conditions 
of the early universe. An exponential phase of the expansion of the universe is generated by a scalar field $\varphi$, the inflaton, with a small mass (compared to the Planck mass) $\mu$. The smallness of the
inflaton mass suggests a pseudo-Nambu-Goldstone origin. Probably the best option proposed in the
literature is a global shift symmetry $\varphi \to \varphi + c$, eventually broken to a discrete subgroup
\cite{freese},\cite{knp}, \cite{kaloper}.  

For trans-Planckian field values the contributions of Planck-scale suppressed higher-\linebreak dimensional operators to the inflationary potential are generically relevant. It is therefore important to consider large-field inflation in the context of some ultraviolet completion, for which string theory is the leading candidate, described by supergravity in its low-energy limit. 
In supergravity, the usual $\eta$-problem can be avoided for a shift symmetric K\"ahler potential   \cite{Kawasaki:2000yn} $K = K(\p+\pb)^2)  \,. $
The invariance is here with respect to $\p \rightarrow \p + i c$, where $c$ is a real constant,  and the inflaton is  $\varphi = \sqrt 2\, \text{Im}\,\phi$. 
 
One of the simplest realization of inflation is of large-field, type, achieved with a free massive scalar field, $ V = \mu^2 \varphi^2 \,$. 
In addition to primordial curvature perturbations, which have been measured with remarkable accuracy \cite{Ade:2013uln}, it predicts sizeable tensor perturbations \cite{Lyth:1996im} for which evidence has been reported recently \cite{Ade:2014xna}.

However, the simplest supersymmetric extension of the potential   $\mu^2 \varphi^2 \,$  defined by the superpotential 
\begin{align}\label{naiv}
W = \frac12 \mu \p^2\,,
\end{align}
has a well-known problem generated by the shift symmetry. Due to the negative term $- 3 |W|^2$ in the supergravity scalar potential, for large values of the inflaton field the potential becomes 
$ V(\varphi) \sim - 3 \mu^2\varphi^4\,$ and the potential is unbounded from below.

The problem can be avoided by introducing an additional `stabilizer field' $S$, which has no shift symmetry in the K\"ahler potential \cite{Kawasaki:2000yn}, i.e.,
\begin{align}\label{eq:chaoticK}
K = K( (\p+\pb)^2 , |S|^2) \ \ ,
\end{align}
together with the superpotential
\begin{align}\label{intro1}
W_{ \rm inf} = \mu S \p \,, 
\end{align}
which breaks the shift symmetry softly. 
This model has been generalized to a class of chaotic inflation models by replacing the inflaton field $\p$ by a function $f(\p)$ in the superpotential \cite{klr}. For recent studies of chaotic inflation in supergravity and string theory, see \cite{sugra} and \cite{string1,string2,Marchesano:2014mla}, respectively. 

Another popular inflationary model is the Starobinsky model \cite{starobinsky}, which has a dual interpretation. On one hand, it is a gravitational theory with a higher-derivative term $R^2$. On the other hand, it can be described as Einstein gravity coupled to a scalar, with a very particular scalar
potential. The model was generalized to supergravity in \cite{cecotti,fkvp}, where it was shown that, in a chiral multiplet formulation, in addition to the inflaton multiplet, there is a second chiral multiplet needed\footnote {It is also possible to realize the Starobinsky model by using massive vector multiplets \cite{porrati}.}. It was subsequently shown in \cite{kl2} that the stabilizer needs additional interactions in order to stabilize its vev to zero during inflation. In the chiral formulation, the second chiral multiplet can also be replaced by a nonlinear superfield, where the corresponding scalar is absent \cite{adfs}. The couplings of this second chiral multiplet to the inflaton are very similar to the
previously discussed case of the stabilizer field in chaotic inflation and could plausibly have a similar microscopic origin. 
  
One of the open questions is the origin of the shift-symmetry breaking in  (\ref{intro1}). It seems unnatural from a string theory viewpoint to mix a field with a shift-symmetry to another field  with no such symmetry. This is  true in particular in flux compactifications, which is a generic framework
invoked for generating such superpotential mass terms.  In the following we propose a natural interpretation of the stabilizer field and of such a coupling in terms of a three-form
multiplet, both from the viewpoint of the soft breaking of the shift symmetry and from string theory. 
More precisely, we will show that the mass term (\ref{intro1})  is  uniquely singled out by requiring the shift symmetry $\phi \to \phi + i c $ and invariance under the three-form gauge symmetry.

The three-form was to our knowledge used in chaotic inflaton for the first time in \cite{kaloper}, which
noticed the nice role of the shift-symmetry in this case,  interpreted it as a "natural inflation" setup.    
It was also discussed recently in \cite{Marchesano:2014mla} in a string theory setup, as a concrete F-term string realization of axion-monodromy \cite{monodromy},\cite{knp}.   
%

\section{Three-form and shift symmetry}

Let us start from a lagrangian containing a scalar $\varphi$ and a three-form field $C_{mnp}$, having
a global shift symmetry $\varphi \to \varphi + c$, up to boundary terms,
\begin{equation}
{\cal S}_0 = \int d^4 x \ \{ - \frac{1}{2} (\partial \varphi)^2 - \frac{1}{2 \times 4 !} F_{mnpq}^2 +
\frac{\mu}{24} \ \varphi \ \epsilon^{mnpq} F_{mnpq} \} \ , \label{shift1}
\end{equation}
where 
\begin{equation}
F_{mnpq} = \partial_m C_{npq} + {\rm 3 \ perm.} \ . \label{shift2}
\end{equation}
For future convenience we define
\begin{equation}
F = \frac{1}{4 !} \epsilon^{mnpq} F_{mnpq} \ , \ F_{mnpq} = - \epsilon^{mnpq} F  \ . \label{shift3}
\end{equation}
 
The lagrangian (\ref{shift1}) has actually to be supplemented with a boundary term 
\begin{equation}
{\cal S}_b = \frac{1}{6} \int d^4 x \ \partial_m \left( F^{mnpq} C_{npq} - \mu \varphi \epsilon^{mnpq}
C_{npq} \right)  \ , \label{shift4}
\end{equation}
in order to find the correct field equations. It is interesting to notice that, whereas the "bulk" action
(\ref{shift1}) has a shift symmetry $\varphi \to \varphi + c$ only up to boundary terms, the total action
\begin{eqnarray}
&& {\cal S}  = {\cal S}_0 +{\cal S}_b =   \int d^4 x \ \{ - \frac{1}{2} (\partial \varphi)^2 - \frac{1}{2 \times 4 !} F_{mnpq}^2 - \frac{\mu}{6} \ \epsilon^{mnpq} \partial_m \varphi \  C_{npq} \} \nonumber \\
&& + \frac{1}{6} \int d^4 x \ \partial_m \left( F^{mnpq} C_{npq}  \right)   \label{shift5}
\end{eqnarray}
is exactly shift symmetric. A massless three-form field has no on-shell degrees of freedom. As such, it
can be integrated out via its field eqs. 
\begin{equation}
\partial^m F_{mnpq} \ = \ - \mu \ \epsilon_{mnpq} \partial^m \varphi  \ , \label{shift6}
\end{equation}
whose solution is given by
\begin{equation}
F = - \mu \varphi + f_0 \ , \label{shift7}
\end{equation}
where $f_0$ is a constant, which is to be interpreted as a flux. It was argued in \cite{bp} that $f_0$
is quantized in units of the fundamental electric coupling $f_0 = n e^2$, fact that was argued to  have important consequences for the landscape of string theory.
After doing so, the final lagrangian takes the form
\begin{equation}
{\cal S} = \int d^4 x \ \{ - \frac{1}{2} (\partial \varphi)^2 - \frac{1}{2} (\mu \varphi-f_0)^2 \} \ . \label{shift8}
\end{equation}
Notice that the boundary term ${\cal S}_b$ is crucial in obtaining the correct action. Ignoring it leads
to the wrong sign of the last term in  (\ref{shift8}), fact that created confusion in the past. 
In the form  (\ref{shift8}), it is clear that the theory describes a massive scalar field of mass m, whereas the flux $f_0$ determines the ground state. It is remarkable that, whereas the action has a shift symmetry that would naively suggests that the scalar is massless, actually the field acquires a topological mass \cite{kaloper}.
In the final formulation (\ref{shift8}) the shift symmetry seems completely broken. There is one
sense in which a discrete subgroup of it is preserved, however, namely
\begin{equation}
\varphi \to \varphi + \frac{e^2}{\mu} \quad, \quad n \to n + 1 \ , \label{shift9}
\end{equation} 
where $n$ is the flux quantum. A nice string intepretation of (\ref{shift9}) in terms of axion monodromy
and brane nucleation  was recently provided in \cite{Marchesano:2014mla}, following \cite{kaloper}.
 
\subsection{Dual formulation}

The dual formulation contains only a massive three-form field. The duality proceeds starting from the
master action
\begin{equation}
{\cal S}_0 = \int d^4 x \ \{  \frac{\mu^2}{2} V_m^2 - \frac{1}{2 \times 4 !} F_{mnpq}^2 -
\frac{\mu}{6} \ \ \epsilon^{mnpq} V_m (H_{npq}-C_{npq}) \} \ , \label{shift10}
\end{equation} 
where $H_3 = d B_2$ is the field strength of a two-form field $B_2$, in form language. Field eqs. of $B$
gives
\begin{equation}
d V = 0 \quad \to V = - d \varphi  \ . \label{shift11}
\end{equation}
Plugging back in  (\ref{shift10}) one finds the original action (\ref{shift1}). Alternatively, eliminating the vector field through its field eqs. leads to
\begin{eqnarray}
&& V^m = \frac{1}{6\mu} \ \epsilon^{mnpq}  (H_{npq}-C_{npq}) \ , \nonumber \\
&& S_{\rm dual} = \int d^4 x \ \{  - \frac{\mu^2}{12} (H_{npq}-C_{npq})^2 -  \frac{1}{2 \times 4 !} F_{mnpq}^2 \} \ . \label{shift12}
\end{eqnarray}
In the dual formulation, the massive three-form has one degree of freedom, matching the degree of freedom of the scalar in the original formulation. In the action (\ref{shift12}) the three-form absorbed the
two form $B_2$ and its axion in a generalization of the Stueckelberg mechanism (see for ex. \cite{quevedo}), which is transparent writing the action in the more compact form
\begin{equation}
{\cal S}_{\rm dual} = \int \ \{  - \frac{\mu^2}{2} (dB_2-C_3) \wedge ^\star (dB-C_3) - \frac{1}{2} F_4  \wedge ^\star F_4 \} \ . \label{shift13}
\end{equation}  
\section{Three-form multiplet in supersymmetry and stabilizer multiplet in inflation}

The three-form multiplet in supersymmetry is defined as the real superfield \cite{gates,threeform,bggp,bgg,gls}
\begin{eqnarray}
&& U = {\bar U} = B + i (\theta \chi - {\bar \theta} {\bar \chi}) + \theta^2 {\bar M} +
 {\bar \theta}^2 { M} + \frac{1}{3} \theta \sigma^m {\bar \theta} \epsilon_{mnpq} C^{npq}
+ \nonumber \\ 
&& \theta^2 {\bar \theta}  (\sqrt{2} {\bar \lambda} + \frac{1}{2} {\bar \sigma}^m \partial_m \chi)
+ {\bar \theta}^2 \theta  (\sqrt{2} {\lambda} - \frac{1}{2} {\sigma}^m \partial_m {\bar \chi})
+ \theta^2 {\bar \theta}^2 (D-\frac{1}{4} \Box B) \ . \label{susy1}
\end{eqnarray}
The difference between $U$ and a regular vector superfield $V$ is the replacement of the vector potential
$V_m$ by the three-form $C^{npq}$. 
In order to find correct kinetic terms, the analog of the chiral field strength superfield $W_{\alpha}$
for a vector multiplet is replaced by the chiral superfield \cite{gates}
\begin{equation}
S = - \frac{1}{4} {\bar D}^2 U \quad , \quad S (y^m,\theta) = M + \sqrt{2} \theta \lambda +
\theta^2 (D + i F) \ , \label{susy2}
\end{equation}
with $F$ defined as in (\ref{shift3}). The definition (\ref{susy2}) is invariant under the gauge transformation $U \to U - L$, where $L$ is a linear multiplet. Correspondingly, lagrangians expressed
as a function of $S$ will have this gauge freedom.   One can therefore choose a gauge in which $B=\chi=0$  in  (\ref{susy1}) and the physical fields are $M,\lambda$. 
The supersymmetrization of the coupling of the inflaton $\phi$ to the three-form is a superpotential mass term coupling a chiral superfield $\phi$ including the inflaton 
\begin{equation}
[\mu S \phi ]_F + {\rm h.c.} = [\mu (\phi + {\bar \phi}) U]_D - {\cal S}_b(\phi) \ , \label{susy3}
\end{equation}
where ${\cal S}_b (\phi)$ is a total  derivative, given explicitly by
\begin{equation}
{\cal S}_b(\phi) = \mu \ \partial^m \left[ \frac{1}{4} \left(B \partial_m (\phi+{\bar \phi}) - 
\partial_m B  (\phi+{\bar \phi})\right) + \frac{1}{2 \sqrt{2}} (\chi \sigma_m {\bar \psi} +
\psi \sigma_m {\bar \chi}) - \frac{i}{6} \epsilon_{mnpq} (\phi-{\bar \phi}) C^{npq} \right] \ . \label{susy03}
\end{equation}
The inflaton $\varphi$ is contained in the imaginary part of the lowest component $\phi_| = (\zeta + i \varphi)/\sqrt{2}$. Notice that in the generalization of the stabilizer models proposed in \cite{klr}, the superpotential $W = f(\phi) S$ can also be re-written
as a contribution to the Kahler potential
\begin{equation}
[S f(\phi)]_F + {\rm h.c.} = [( f(\phi) + {\bar f} ({\bar \phi})) U]_D - {\cal S}_b (f(\phi))\ , \label{susy031}
\end{equation}
where ${\cal S}_b (f(\phi))$ is a boundary term generalizing (\ref{susy03}) that we don't display
here. 
However only for the linear case $f(\phi) = \mu \phi$ is the shift symmetry unbroken in the action, up to boundary terms. On the other hand, shift symmetry is preserved by additional terms in the Kahler potential of the type $[(\phi+{\bar \phi}) g(U)]_D$. However, only for a linear function $g(U) = U$
is this term invariant under gauge transformations of the three-form $U \to U - L$.  The linear coupling
$\mu [(\phi+{\bar \phi}) U]_D = [\mu \phi S]_F + {\rm h.c.} + {\rm total \ deriv.}$ is therefore uniquely singled out by requiring shift symmetry and three-form gauge symmetry. 

Let us consider the simplest\footnote{We will comment later on expected changes by considering a more general Kahler potential.} example of interest for applications to inflation, provided by the lagrangian containing the chiral superfields $S$ and $\phi$
\begin{eqnarray}
&&K = |S|^2 + \frac{1}{2} (\phi+{\bar \phi})^2 + {\cal S}_b\ , \nonumber \\
&& W = \mu S \phi \ . \label{susy4}
\end{eqnarray} 
where the boundary action ${\cal S}_b$ is given by
\begin{eqnarray}
&& {\cal S}_b = {\cal S}_b (\phi) + {\cal S}_b (C) \nonumber , \\
&& {\cal S}_b (\phi) \ = \ \mu \int d^4 \theta \ (\phi + {\bar \phi}) U - \mu (\int d^2 \theta \ S \phi 
+ {\rm h.c.} ) \ , \nonumber \\
&&  {\cal S}_b (C) = \frac{1}{2} \left[ D^{\alpha} (S D_{\alpha} U - U D_{\alpha}S  ) +{\rm h.c.}  \right] 
\label{susy04}
\end{eqnarray}
and is needed, as in the previous section, in order to get consistent field eqs. The shift symmetric term of the kinetic term $\frac{1}{2} (\phi+{\bar \phi})^2$ is equivalent in the global supersymmetry case, up to boundary terms which are innocent (unlike the ones containing the three-form), to the standard one $|\phi|^2$. We keep however this form for later generalizations to supergravity. The full lagrangian can also be written as only a contribution to the Kahler potential
\begin{equation}
K =  |S|^2 + \frac{1}{2} (\phi+{\bar \phi})^2 + \mu (\phi + {\bar \phi}) U + {\cal S}_b (C)
\ . \label{susy5} 
\end{equation}
Notice that in the form (\ref{susy5}) the coupling inflaton-three form is precisely in the form 
(\ref{shift5}), which includes therefore the inflaton-dependent boundary term  ${\cal S}_b (\phi)$, as seen from the explicit expression
\begin{eqnarray}
&& \left[(\phi+{\bar \phi}) U \right]_D =  F_{\phi} M + {\bar F}_{\bar \phi} {\bar M}
+ D (\phi + {\bar \phi}) - \frac{i}{6} \epsilon_{mnpq} \partial^m (\phi-{\bar \phi}) C^{npq}
- (\lambda \psi + {\bar \lambda} {\bar \psi}) +   \nonumber \\
&& + \partial^m \left[ \frac{1}{4} \left(B \partial_m (\phi+{\bar \phi}) - \partial_m B  (\phi+{\bar \phi})\right) + \frac{1}{2 \sqrt{2}} (\chi \sigma_m {\bar \psi} +
\psi \sigma_m {\bar \chi}) \right] \ . \label{susy05}
\end{eqnarray}
Field equations determine the auxiliary fields to be
\begin{eqnarray}
&& 2 D + \mu (\phi + {\bar \phi}) = 0 \quad , \quad F =  -\frac{i\mu}{2} (\phi - {\bar \phi}) - f_0 \ , \nonumber \\
&& F_{\phi} + \mu {\bar S} = 0 \ , \label{susy05} 
\end{eqnarray}
where $f_0$ is a flux allowed since $F$ is a field strength and not really an auxiliary field, such that
its field eq. is  $ \partial_{n }(F  +\frac{i\mu}{2} (\phi - {\bar \phi})) = 0$. 
The final lagrangian is obtained after taking into account carefully the contribution of the boundary terms. The scalar potential is given by
\begin{equation}
V = \mu^2 |M|^2 + |\mu \phi + i f_0|^2 \ , \label{susy06} 
\end{equation}
and display again the combination inflaton/flux similar to (\ref{shift8}). As already discussed in
in the non-supersymmetric case and displayed in (\ref{shift9}), the shift symmetry is broken to a discrete subgroup, with a corresponding change in the flux quantum. 

Notice that for the purpose of finding the correct on-shell lagrangian and scalar potential, there is a simpler formulation in which $S$ is treated as a standard chiral superfield with $D+iF$ as standard
auxiliary fields, no boundary terms are included, but the superpotential of the theory is changed
according to \cite{bggp,bgg,gls}
\begin{equation}
W (\phi, S) \to W' (\phi, S) \ = \ W (\phi, S) + i f_0 S \ , \label{susy07} 
\end{equation}     
which in our case becomes 
\begin{equation}
W' = \mu S \phi + i f_0 S \ . \label{susy08} 
\end{equation}
Similarly to (\ref{shift9}), for quantized flux $f_0 = n e^2$, there is a discrete remnant of the shift symmetry
\begin{equation}
\phi \to \phi - \frac{i e^2}{\mu} \quad , \quad n \to n + 1 \ ,   \label{susy09}  
\end{equation}
interpreted in terms of membrane nucleation which induces the monodromy in the inflaton excursions.  
 
\subsection{Dual formulation}

The dual formulation starts from the master action
\begin{equation}
K = |S|^2 + \frac{\mu^2}{2} V^2 + \mu^2 V (U-L)  \ , \label{susy6}
\end{equation}
 where $V$ is a real vector superfield and $L$ is a linear multiplet satisfying $D^2 L = {\bar D}^2 L =0$, that can be expressed as a function of the unconstrained fermionic
 superfield $\Sigma_{\alpha}$ via $L = D^{\alpha} {\bar D}^2 \Sigma_{\alpha} +
 {\bar D}_{\dot \alpha} {D}^2 {\bar \Sigma}^{\dot \alpha}  $. Field eq. of the linear multiplet gives
\begin{equation}
\mu V \ = \ \phi + {\bar \phi}  \ , \label{susy7}
\end{equation} 
which, after plugging back into (\ref{susy6}), gives the original bulk Kahler potential (\ref{susy5}).
On the other hand, eliminating the vector superfield via its field eqs. leads to
\begin{equation}
V = -U + L \quad , \quad K_{\rm dual} = |S|^2 - \frac{\mu^2}{2}(U-L)^2 \ . \label{susy8}
\end{equation}
The dual lagrangian  (\ref{susy8}) contains a massive three-form multiplet, which has precisely the same degrees of freedom as the original action containing two chiral superfields $S$ and $\phi$. 
Notice that the combination $U-L$ is the analog of the Stueckelberg combination $V - d \phi$ for a massive vector multiplet and is gauge invariant in the same sense. In the massive case, all bosonic
$B,M, C_{mnp}$ and fermionic fields $\chi,\lambda$ are physical. The action  (\ref{susy8}) contains
therefore four bosonic and four fermionic degrees of freedom, of mass $\mu$. They match of course the degrees of freedom of the chiral multiplet formulation in terms of the chiral fields $\phi,S$.  Interestingly enough, in analogy with the non-supersymmetric starting point, the massive three-form multiplet contains both the inflaton and the stabilizer field, and its mass term drives chaotic inflation. 
\subsection{Corrections to the inflaton potential}

Whereas the mass term (\ref{susy4}) or equivalently the D-density  $[(\phi+ {\bar \phi}) U]_D$ is uniquely singled out by the shift symmetry and  the three-form gauge symmetry, more general Kahler (or higher derivative) contributions can be considered. As shown in \cite{kl2}, corrections to the stabilizer
Kahler potential, for ex. a term of the type $- \zeta ({\bar S} S)^2$ are actually needed in order
to generate a large stabilizer mass during inflation, without changing the inflaton potential. A more general Kahler potential of the form $K (S,{\bar S}, \phi+{\bar \phi})$ does not change conceptually our discussion above provided it contains in its expansion the standard quadratic terms, and does not impact inflationary dynamics provided that its stabilizes the field $M$ to zero during inflation.\\
On the other hand, corrections to the inflaton potential arise from higher-derivative interactions. The simplest higher-derivative ghost-free correction to the effective action is of the form
\begin{equation}
\frac{1}{\Lambda^4} \ [D^{\alpha} S D_{\alpha} S \ D_{\dot \alpha} {\bar S} D^{\dot \alpha} {\bar S}]_D \ , 
\label{corr1} 
\end{equation}
where $\Lambda$ is an UV scale. This generates corrections of the type $\frac{1}{\Lambda^4} F^4$ to the effective action, of the type considered in the non-supersymmetric case in \cite{kaloper}, which in this case lead to corrections to the inflaton potential $\delta V \sim \frac{\mu^4 \varphi^4}{\Lambda^4}$. According to \cite{kaloper}, such corrections to not affect significantly chaotic inflation provided that $\Lambda >> M_{GUT}$.      
\section{Supergravity formulation of the three-form multiplet}

The supergravity embedding of the three-form multiplet was pioneered in \cite{threeform,bggp,bgg,ow}. In what follows we use the notations and conventions of \cite{fvp}. The chiral weight $c$of the three-form
multiplet $U$ in supergravity is zero, because it is real. The Weyl weight $w $, on the other hand, is arbitrary. It is convenient to take it equal to $2$, so in what follows  $(c_U,w_U)=(0,2)$. We also define the chiral projector $\Sigma$, of weights  $(c_{\Sigma},w_{\Sigma})=(3,1)$. In the old minimal supergravity, the compensator $S_0$ has weights $(c_0,w_0)=(1,1)$ and it is fixed at $S_0 = {\bar S}_0
= e^{K/6}$ in order to define supergravity in the Einstein frame. All other chiral fields are
defined in order to have zero chiral and Weyl weights. One can then define the analog of the chiral superfield $S$ in the previous section by
\begin{equation}
S \ = \ \frac{1}{S_0^3} \ \Sigma (U) = \  e^{- \frac{K}{2}} \ \Sigma (U) \ . \label{sugra1}
\end{equation}
The inflaton will be one of the matter fields with zero weights, such that an arbitrary supergravity
lagrangian will be of the form
\begin{equation}
{\cal S} = \left[ - 3 e^{- \frac{1}{3} K (\phi,S,{\bar \phi},{\bar S})} S_0 {\bar S}_0 \right]_D + \left[ S_0^3 W (\phi,S) \right]_F
\ . \label{sugra2}
\end{equation} 
Of particular interest in what follows for inflationary models is the mass-like term
\begin{equation}
\left[ \mu S_0^3 \phi S \right]_F = \left[ \mu (\phi + {\bar \phi}) U \right]_D -{\cal S}_b \ , \label{sugra3}
\end{equation}
where ${\cal S}_b$ is a boundary term.  
In analogy with the rigid limit therefore, the would-be mass term does not  break the shift symmetry. The boundary term is expected, similar to the rigid case, to be completely included in the
lagrangian with  the term  $\mu (\phi + {\bar \phi}) U$ in the Kahler potential.

It was shown in \cite{bggp,bgg} that, similarly to the global supersymmetry case,  the supergravity couplings of the three-form can be described by using the chiral superfield $S$, treated as a standard
chiral superfield, with the modification (\ref{susy07}) of the superpotential. This is the simplest approach that we will use in what follows.    
\subsection{Chaotic inflation}
The lagrangian for chaotic inflation is provided by \cite{Kawasaki:2000yn}
\begin{eqnarray}
&& K \ = \ |S^2| +  \frac{1}{2} (\phi+{\bar \phi})^2 \nonumber , \\ 
&& W \ = \ \mu S \phi + i f_0 S \ , \label{sugra4} 
\end{eqnarray}
where we added the flux contribution $f_0$ for practical calculations (allowing to compute naively scalar potential and field eqs.) and we neglected Kahler corrections for the stabilizer \cite{kl2}, which are important for the inflationary dynamics but not for our current discussion. 
We have shown in Section 3 that in the global supersymmetry case there is a dual formulation in terms
of a massive three-form multiplet, of lagrangian (\ref{susy8}). The supergravity generalization is similar. Starting from the master action
\begin{equation}
{\cal S} = \left[ - 3 e^{- \frac{1}{3} K (S,{\bar S},\mu V)} S_0 {\bar S}_0 + \mu^2 V (U-L) \right]_D 
\ , \label{sugra5}
\end{equation}
where $V$ is a vector multiplet of weights $(c_V,w_V)= (0,0)$ and $L$ is a linear multiplet of the same weights as the three-form multiplet $U$, $(c_L,w_L)= (0,2)$, eqs. of motion of the linear multiplet and the action can be written as
\begin{eqnarray}
&&  \mu V = \phi + {\bar \phi} \nonumber , \\
&&  {\cal S} = \left[ - 3 e^{- \frac{1}{3} K (S,{\bar S}, \phi + {\bar \phi} )} S_0 {\bar S}_0 + 
\mu (\phi + {\bar \phi})  U \right]_D 
\ .  \label{sugra6}
\end{eqnarray}
In the dual version, one uses the field eqs. of the vector multiplet
\begin{eqnarray}
 \frac{\delta K }{\delta V} e^{- \frac{1}{3} K} S_0 {\bar S}_0 + \mu^2 (U-L) \ = \ 0 \ .  \label{sugra7}
\end{eqnarray}
For example for a quadratic form $K = \frac{\mu^2}{2} V^2$, after eliminating the vector multiplet
$ V = - U + L$, we recover the mass term of a massive three-form multiplet (\ref{susy8}).
\subsection{The Starobinsky model}

The equivalence between higher-derivative supergravity and standard supergravity with two chiral superfields was pioneered by Cecotti \cite{cecotti} and developed further in \cite{fkvp}. In what follows, we discuss the Starobinsky model and duality in the case where one of the chiral fields contain the three form.  
In the simplest, chiral formulation, the Starobinsky model is given by 
\begin{eqnarray}
&& K \ = \ - 3 \ln (T + {\bar T} - |S^2|)  \nonumber , \\ 
&& W \ = \ \mu (T-\frac{1}{2}) S + i f_0 S \ , \label{sugra8} 
\end{eqnarray}
where again $S$ is the chiral superfield containing the four-form field strength, but treated as a standard chiral superfield in  (\ref{sugra8}), due to the addition of the flux superpotential linear term proportional to $f_0$. The inflaton $\varphi$ is defined here via the real part of $T$: 
\begin{equation}
T \ = \ e^{\sqrt{\frac{2}{3}} \varphi} + i \sqrt{\frac{2}{3}} \ a \ , \label{sugra9} 
\end{equation}
with $a$ being an axion.  Similar to the previous cases, the model has a discrete shift symmetry acting on the axion field. If the flux is quantized $f_0 = n e^2$, with $e$ being the elementary three-form electric charge and with our definition (\ref{sugra9}), the symmetry transformation is
\begin{equation}
a \to a - \sqrt{\frac{3}{2}} \frac{e^2}{\mu} \quad , \quad  n \to n+1 \  \label{sugra09} 
\end{equation}
but it does not involve the inflaton $\varphi$. This version of Starobinsky model cannot therefore be
considered as "natural" in the sense that super-Planckian values of the inflaton cannot be reached
by nucleating three-form membranes.    \\

- {\bf Dual gravitational formulation} \\

The dual description starts from the lagrangian
\begin{eqnarray}
&& {\cal S} = - \left[  (1+ T + {\bar T} - |S|^2) S_0 {\bar S}_0 \right]_D +
\left[ (\mu T + i f_0) S S_0^3 \right]_F = \nonumber \\
&&  - \left[  (1 - |S|^2) S_0 {\bar S}_0 \right]_D +
\left[ T (\mu S - \frac{{\cal R}}{S_0}) S_0^3 + i f_0 S S_0^3 \right]_F   \ , \label{sugra10} 
\end{eqnarray} 
where in  (\ref{sugra10}) ${\cal R}$ denotes the chiral gravity multiplet superfield and in the last equality we used the identity \cite{fvp,fkvp}
\begin{equation}
\left[ (f(\Lambda) + {\bar f} ({\bar \Lambda})) S_0 {\bar S}_0 \right]_D =
\left[ f(\Lambda) {\cal R} S_0^2  \right]_F  + {\rm total \ derivative} \ . \label{sugra11} 
\end{equation} 
One can therefore eliminate the chiral multiplet $S$ in favor of the gravity multiplet ${\cal R}$, according to
\begin{equation}
{\cal R} = \mu S_0 S = \frac{\mu}{S_0^2} \Sigma (U) = \frac{1}{S_0} \Sigma ({\bar S}_0) 
\ , \label{sugra12}  
\end{equation}
where the last equality defines actually the chiral  multiplet ${\cal R}$ in supergravity. 
In detail, the components of the chiral superfield ${\cal R}$ are 
\begin{equation}
{\cal R} = \left( {\overline u} \equiv S \,+\, i P\, , \ \gamma^{mn} {\cal D}_m \psi_n \, ,\ - \ \frac{1}{2}\ R
\,-\, \frac{1}{3} \ A_m^2 \,+\, i \,{\cal D}^m A_m \,- \,\frac{1}{3} \ u \, {\overline u}  \right)  \ , \label{staro01}
\end{equation}
where $u$ and $A_m$ are the ``old minimal'' auxiliary fields of $N=1$ supergravity and $\psi_n$ is the gravitino field.  Notice that according to the prescription  (\ref{susy07}), if the flux term $f_0$ is included as in (\ref{sugra10}), $S$ can be treated as a standard chiral multiplet in  the lagrangian. 
Let us however ignore this term and look at the duality with $S$ containing the four-form field strength. 
In components, (\ref{sugra12}) contains the duality relations
\begin{eqnarray}
&& {\bar u} = \mu M \quad , \quad \gamma^{mn} {\cal D}_m \psi_n = \sqrt{2} \ \mu \lambda \ , \nonumber \\
&& - \frac{1}{2} R - \frac{1}{3} A_m^2 - \frac{1}{3} |u^2| = \mu D  \quad , \quad
{\cal D}^m A_m = \mu F \ . \label{sugra011}
\end{eqnarray}
In the dual formulation, the vector multiplet auxiliary field $A_m$ of the
old minimal supergravity is therefore replaced by the three-form $C_{mnp}$ and the duality relation 
(\ref{sugra011}) contain the duality $C_3 = ^\star A$, or in components $C_{mnp} = \epsilon_{mnpq}
A^q$.  Duality  (\ref{sugra12}) also implies the  relation 
$\mu (U-L) = S_0 {\bar S}_0 $. 
It is unclear to us if this could be interpreted as replacing the chiral compensator $S_0$ of the old minimal supergravity by a three-form
compensator\footnote{A different formulation of supergravity with a three-form compensator multiplet  was  proposed some time ago in \cite{ow}.}. \\
For the dual lagrangian, the simplest option is probably to add the flux superpotential term $f_0$ 
as in (\ref{sugra10}) and treat $S$ as a standard chiral superfield.  One therefore finds, in the old minimal supergravity formulation 
\begin{equation}
{\cal S}_{\rm dual} = - \left[ S_0 {\bar S}_0 - \frac{1}{\mu^2} {\cal R}
{\bar {\cal R}}  \right]_D + \left[ i f_0  \frac{\cal R}{\mu} S_0^2 \right]_F \ . \label{sugra13} 
\end{equation}
According to (\ref{sugra11}), the last, new term compared to the standard higher-derivative supergravity in (\ref{sugra13}), is a total derivative. 
\section{Embedding in string theory}

A natural interpretation of the inflaton in string theory is as a Wilson line \cite{string1,Marchesano:2014mla}, which being an internal component of a gauge field, it enjoys the shift symmetry as a remnant of the higher-dimensional gauge symmetry.  On the other hand, the three-form
origin could be one of the RR forms present in the closed string spectrum of type II strings.
Let us give a suggestive example of one D5 brane in type IIB strings, without getting into various possible subtleties; it is by no means to be considered as a unique possibility. There is a $U(1)$
gauge field living on the brane. In what follows we denote by $x_5,x_6$ the internal dimensions in
the brane wordvolume and by $A_5,A_6$ the internal component of the gauge fields. After a suitable complexification (we take the complex structure of the torus $\tau=i$ to simplify the discussion) 
\begin{equation}
z = \frac{x_5+i x_6}{\sqrt{2}} \quad , \quad  \phi = \frac{A_6+i A_5}{\sqrt{2}} \ , \label{string1} 
\end{equation}
the one-form gauge field and field strengths are
\begin{equation}
A = A_M d x^M = A_m d x^m  + 2 \ {\rm Im} \ (\phi dz ) \quad , \quad   
F= \frac{1}{2} F_{mn} d x^m \wedge dx^n  + 2 \ {\rm Im} \ (d \phi \wedge dz )  \ . \label{string2}
\end{equation}
Type IIB strings have a four-form that can contain three-forms from the four-dimensional viewpoint
\begin{equation}
C_4 \ \supset \ C_3^z dz + {\bar C}_3^z d{\bar z} \ , \label{string3}
\end{equation}
that in components read $C_3^z = \frac{1}{\sqrt{2} \times 3 !} (C_{mnp5}-i C_{mnp6}) dx^m \wedge dx^n \wedge dx^p$. 
Then the Chern-Simons couplings of the RR forms to the brane worldvolume gauge field is given by
\begin{equation}
q_5 \int_{D5} C \wedge e^F  \supset \ \mu \int (d\phi \wedge {\bar C}_3^z + d {\bar \phi} \wedge {C}_3^z) 
= - \mu \int (\phi \wedge {\bar F}_4^z +  {\bar \phi} \wedge {F}_4^z) + {\rm total \ deriv.}  \ , \label{string4}
\end{equation}
where $q_5$ is the D5 brane RR charge and $\mu = q_5 \int_{{\cal C}_2} {\Phi} {\bar {\cal C}}_3^z$, where
${\cal C}_2$ is the two-cycle wrapped by the brane and ${\Phi}, {\bar {\cal C}}_3^z$ are the internal profiles of the corresponding fields. The flux parameter $f_0$ of the previous sections is related by Hodge duality to the five-form flux along the internal space. 
One concrete setup is the orientifold of type IIB string
by $\Omega' = \Omega {\cal I}_4$, where ${\cal I}_4$ is the inversion of four (two complex) internal  coordinates $z_1,z_2$. The RR 4-form $C_4$ is odd under $\Omega$, but its components ${C}_3^{z_1},  {C}_3^{z_2} \sim \int_{{\cal C}_1} C_4$, integrals over one-cycles in the $z_1,z_2$ internal space are even. The D5 brane under consideration  should wrap $z_1$ or $z_2$. In this case,
the inflaton mass parameter $\mu$ is determined by the D5 brane charge $q_5$, and also by the wavefunction normalization of the Wilson line kinetic term, which depends in general on complex structure moduli. A small  value $\mu \sim 10^{-5} M_P$ could then be obtained for extreme values of complex structure moduli. 

Another possible realization is the type I string with magnetized \cite{bachas} D9 branes \footnote{In what follows $m_1 \cdots
m_4$ are spacetime indices, whereas $i_1 \cdots i_6$ are internal indices.}. In this case, the Chern-Simons couplings of RR fields to D9 brane gauge fields are
\begin{equation}
q_9 \int_{D9} C \wedge e^F \supset q_9 \int_{D9} C_6 \wedge F \wedge F \  \supset q_9  \int_{D9} \ \epsilon^{m_1 \cdots m_4} \epsilon^{i_1 \cdots i_6} 
C_{m_1m_2m_3i_1i_2i_3} \partial_{m_4} A_{i_4} \left\langle F_{i_5i_6} \right\rangle \ , 
\label{string5}
\end{equation}
where  $\left\langle F_{i_5i_6} \right\rangle$ is a magnetic flux and where $C_6$ is the RR six-form, dual to the RR two-form of type I string. Here the inflaton mass $\mu$ is given by an integral over the compact space of  internal wavefunctions and the magnetic flux $\mu \sim q_9 \int {\cal C} {d \Phi} \left\langle F \right\rangle$. The coupling  (\ref{string5}) has a form similar to (\ref{string4}) with appropriate identification of fields. The flux parameter $f_0$ is here related to a seven form field strength flux or, by Hodge duality, by an internal three-form flux $^\star F_7 = F_ 3 = f_0 \Omega$, where $\Omega$ is the holomorphic $(3,0)$ form. 
 Other string theory examples are discussed in \cite{Marchesano:2014mla} (see also \cite{shlaer}). The importance of integrating out consistently four-dimensional three-forms in superstring compactifications was emphasized 
in \cite{grimm}. 

Our discussion here only concerns the origin of the inflaton-stabilizer coupling. In a realistic string setup, other issues have to be addressed, like moduli stabilization and supersymmetry breaking (see for ex. \cite{string1,string2,Marchesano:2014mla}). They are however beyond the goals of this paper. 

\subsection*{Acknowledgments}

I would like to thank W. Buchmuller, S. Ferrara, R. Kallosh and J. Louis  for very useful discussions and correspondence. 
This work has been supported in part by the ERC advanced grant "MassTeV" and the Alexander von Humboldt foundation. Special thanks are due to the Theory Group in DESY-Hamburg for hospitality during the whole project and to MITP-Mainz for hospitality during the last stages of the work. 


\end{document}